# Secure Clustering in DSN with Key Pre-distribution and WCDS

Al-Sakib Khan Pathan and Choong Seon Hong, *Member, IEEE*

*Abstract*—This paper proposes an efficient approach of secure clustering in distributed sensor networks. The clusters or groups in the network are formed based on offline rank assignment and predistribution of secret keys. Our approach uses the concept of weakly connected dominating set (WCDS) to reduce the number of cluster-heads in the network. The formation of clusters in the network is secured as the secret keys are distributed and used in an efficient way to resist the inclusion of any hostile entity in the clusters. Along with the description of our approach, we present an analysis and comparison of our approach with other schemes. We also mention the limitations of our approach considering the practical implementation of the sensor networks.

## I. INTRODUCTION

THE types of services expected from Wireless Sensor Networks demand the inclusion of security for the trustworthiness of the reported data. While the routing of data and information throughout the network should include some sorts of security mechanisms, the initial formation of the network should also be secured to elude any harmful attempts by the potential adversaries. Hence, it is necessary to ensure security from the very beginning-state of the network's operation. This is more applicable for Distributed Sensor Networks as these are envisaged to operate in the presence of adverse or enemy units. A Distributed Sensor Network (DSN) is basically a wireless sensor network with a large number of sensors and large coverage area. It differs from the traditional wireless sensor network in the sense that, it contains considerably huge number of sensors which are intended to be deployed over hostile and hazardous areas where the communications among the sensors could be monitored, the sensors are under constant threat of being captured by the enemy or manipulated by the adversaries. DSN is dynamic in nature in the sense that, new sensors could be added or deleted whenever necessary [1]. These types of networks are suitable for covering large areas for monitoring, target tracking, surveillance and moving object detection which are very crucial tasks in many military and public-oriented operations.

In this paper, we propose an efficient approach for key-predistribution among the sensors which helps for offline rank assignments of the sensors and eventually plays the crucial role to form a network-wide weakly connected dominating set. Our target is to ensure security and to minimize the number of cluster heads while forming the clusters in the network, so that a relatively small set of cluster heads can securely cover the whole network. We assume that, the base station is fully secure and the adversary cannot affect the base station in any way. The scope of this paper is restricted to the secure clustering of the distributed sensor networks. Our later analysis and simulation show that, our approach could perform well to form secure clusters in a distributed sensor network with a minimum number of cluster heads.

The rest of the paper is organized as follows: Section II outlines the related works, Section III presents our model, Section IV proposes our approach and the method for key-predistribution, Section V contains the performance analysis and comparison, and Section VI concludes the paper.

## II. RELATED WORKS

A cluster is a subset of the total set of sensors in a network which might have at least one cluster head capable of manipulating sensed data locally and then sending the gist of that to the base station. Grouping nodes into clusters is a good idea as it helps to divide the network into several separate but interrelated regions. It also helps for efficient routing within the network. Some of the previous works addressed the issue of clustering or group formation in sensor networks. Most of the previous works on clustering of wireless sensor networks do not address the security issues or consider secure environment for the bootstrapping of the whole network. We argue that, this is in most of the cases might not be possible. For example, if sensor networks are profoundly used in the military reconnaissance scenario, both sides might have the technology and while forming the friendly network there could be a hidden and active enemy-sensor network. If the friendly network is to be formed later than the enemy network in a particular area, the hostile sensors might actively try to be included in the clustering process or could try to hinder the formation of any other network within the region. Here we mention some of the works related to the clustering in wireless

Manuscript received May 29, 2006. This work was supported by the MIC and ITRC projects. Dr. C. S. Hong is the corresponding author.

Al-Sakib Khan Pathan is a graduate student and research assistant in the Networking Lab, Kyung Hee University, South Korea (phone: +82 31 201-2987; fax: +82 31 204-9082; e-mail: spathan@networking.khu.ac.kr).

Dr. Choong Seon Hong is a professor in the Department of Computer Engineering, Kyung Hee University, South Korea (phone: +82 31 201-2532; fax: +82 31 204-9082;e-mail: cshong@khu.ac.kr).

sensor networks. [2] presents a distributed expectation-maximization (EM) algorithm suitable for clustering and density estimation in sensor networks. Energy-Aware clustering is addressed in [3], [4], [5], [6], [7] etc. In [17] the authors propose a load-balanced clustering scheme which increases the lifetime of the network. Other works on clustering in sensor networks are [8], [9], [10] etc.

Though we focus on the clustering or formation of the sensor network, our work differs from all of the above as we model our network to form the groups based on offline rank assignments by pre-distribution of keys and using the notion of weakly connected dominating set considering the whole distributed sensor network as a graph. The details of key generation and selection, prevention of DoS attack caused by jamming [16] and after-formation secure data transmissions are beyond the scope of this paper and will be presented in our future publications.

III. OUR MODEL

We consider the topology of the whole distributed sensor network as a unit-disk graph (UDG) [11], G = (V,E), where V is the set of sensors (vertices) in the network and E is the set of direct communication links (edges) between any two sensors.

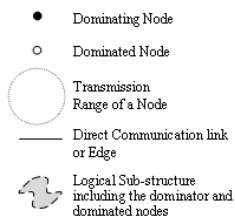 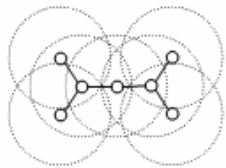

Fig. 1. (Left) Legend used in the paper (Right) Unit-disk Graph

**Definition 1.** A dominating set $S$ is a subset of the vertex set $V$ of a graph G =(V,E) (i.e., $S \subseteq V$ ), so that all other vertices in the graph are adjacent to the vertices of $S$. For a dominating set $S$, $N_G[S]=V$, where $N_G[S]$ is the set of vertices including the vertices in $S$ and the vertices adjacent to a vertex of $S$ (Figure 2). However, finding a minimum size dominating set in a general graph is NP-complete [12].

**Definition 2.** A connected dominating set (CDS), $S_C$ is a dominating set of a given graph G=(V,E) where the induced subgraph of $S_C$ is connected. Figure 3 shows the connected dominating set for our graph model (i.e., all the black vertices).

The connected dominating set for any type of ad hoc network could be used for efficient routing or message transmission throughout the network. However, for CDS, a large number of dominating nodes is needed to maintain the connectivity requirements of the network.

**Definition 3.** A weakly connected dominating set (WCDS), $S_W$ is a dominating set where the graph induced by the stars of the vertices in $S_W$ is connected. A star of a vertex is comprised of the vertex itself and all the vertices adjacent to it (All the black nodes in Figure 4). For any given graph,

$$|WCDS| \leq |CDS| \quad (1)$$

where, |.| denotes the size of the set. So, in case of *WCDS*, less number of dominating nodes is needed for establishing network-wide connectivity than that is required for *CDS*. For example, in Figure 3, $|CDS|$=13 while in Figure 4, $|WCDS|$=8 for the same network size and structure.

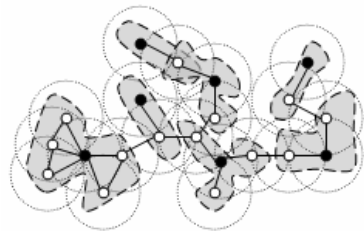

Fig. 2. Dominating Set consisting of black vertices

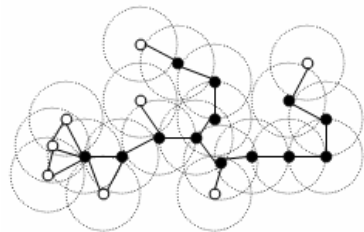

Fig. 3. Connected Dominating Set

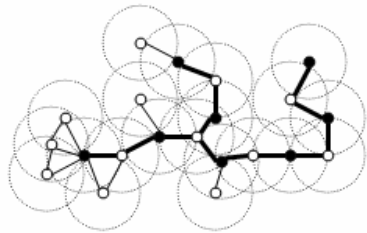

Fig. 4. Weakly Connected Dominating Set

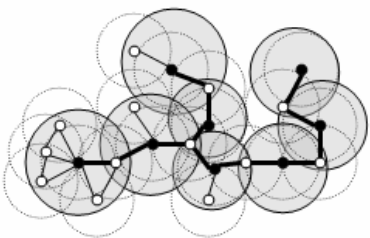

Fig. 5. Dominators' coverage areas in WCDS

The weakly connected dominating set underpins our proposed scheme. In fact, it is easy to see that each dominating node (or vertex) in the weakly connected dominating set is at the center of a star (or, disk). Thus for each dominating node in a *WCDS* of the overall network, we have one star where all the other nodes in the star are just one hop apart (Figure 5). Also it could be observed that, between two stars there is at least one common dominated node which could be used for the communication purpose between two separate stars. We term this common dominated node between two individual stars as 'Mediator' (Figure 6).

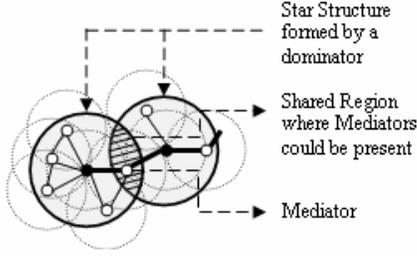

Fig. 6. Mediator between two groups/stars

## IV. OUR APPROACH

We apply two stage operations for secure formation of clusters in the network.

**Assumption 1.** Once the sensors are deployed they remain relatively static in their respective positions.

**Assumption 2.** In a unit disk or transmission range of a sensor, all the neighboring sensors do not necessarily have a direct communication link among themselves. If two nodes *i* and *j* have a direct communication link, it is bidirectional; $\forall_{i,j}, (i,j) \in E \Rightarrow (j,i) \in E$ and it exists if and only if *i* and *j* have common secret keys.

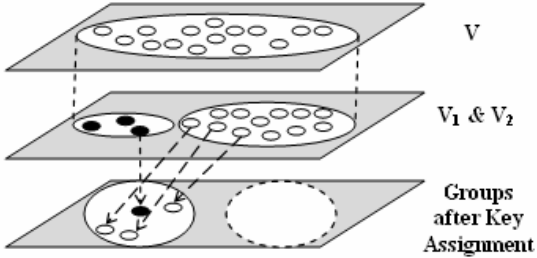

Fig. 7. Ranking of the sensors based on the key pre-distribution

### A. Offline Rank Assignment

The sensors in the network are assigned their ranks based on the offline key-distribution. We divide the whole set of sensors $V$ into two subsets, $V_1$ and $V_2$, where $V_1$ contains the probable group dominators (GD or cluster heads) and $V_2$ consists of ordinary sensors (Os). The set $V_2$ is further divided into several subsets $w_i \subset V_2$, i=1, 2, 3,....,$N$ and $N$ is the maximum number of possible proper subsets of $V_2$. Each $w_i$ is assigned to one element in the set $V_1$. The sensors in the subset $w_i$ (Os$_1$, Os$_2$,....Os$_\eta$) and corresponding one sensor from $V_1$ (let, GD$_i$, i=1) are taken for group-wise key-predistribution (Figure 7). All the sensors in the set $w_i$ are assigned two keys: group key and individual key. The group key is common for all the Oss in a $w_i$ but the individual key is shared by the particular sensor and the GD$_i$. The GD$_i$ contains all the individual keys of the sensors in its $w_i$ and its own group key.

**Assumption 3.** All the sensors have same transmission range. Each node transmits within the transmission range isotropically (in all directions) so that each message sent is a local broadcast.

**Assumption 4.** The Base Station (BS) contains all the individual keys and group keys of the network.

**Assumption 5.** The number of Oss (value of η) in each group is decided on demand. It could be group specific or set to a common value for all the groups. η is actually the maximum degree ($\Delta(GD_i)$) of a GD in a group.

### B. Secure Cluster Formation

The groups of sensors are deployed over the target region one group at a time. After deployment, each Os tries to find out its own GD by sending a join request packet encrypted with its individual key. The corresponding GD in turn sends the join approval message encrypted with the group key. In both cases, both the GD and the Os can decrypt the messages and form the group. In some cases, the corresponding GD of an Os might not be within one-hop transmission range (disk). In this case, the Os detects the presence of other GDs of other groups in its surroundings, collects their ids and sends an error message to the base station (BS) with this information. The GDs within its one-hop transmission range also could detect such erroneous Os and reports to the BS. The BS in turn assigns one of the neighboring GD as the adopter of the orphan Os. In the worst case, the Os might not find any GD in its surroundings. In this case, The BS assigns the rank of a GD (let us call it GD$_{os}$) to that particular Os though it does not contain any other sub-ordinate sensors. An Os which gets its own GD and another GD of another group in its transmission range is the mediator in this case. As stated earlier, all the stars thus shaped could use mediators for the inter-group (inter-star or inter-cluster) communication (see Figure 6). In this way, eventually the resultant logical model of the whole network contains a weakly connected dominating set where the GDs of the logical groups (stars) are the dominating nodes and all other nodes in the network are dominated. This logical model now could be used for secure message delivery within the network (using the secret keys).The pseudo code for secure cluster (group) formation algorithm is presented in Figure 8.

```
Let,
enc_i(.) - message encrypted by individual key of i
enc_iNOT(.) - message encrypted by an unknown
individual key
enc_G(.) - message encrypted by the group key
enc_GNOT(.) - message encrypted by an unknown group key
Os_g - set of Oss allowed under a group dominator g
locbr(.) - local broadcast within one hop
transmission range

for each s ∈ V_Os
      locbr(enc_i(JOIN_REQ))
  if enc_G(JOIN_APRV) from any g ∈ V_GD and hop(s,g)=1
    edge(s,g)
    dominator(s)←g
  else
    flood(enc_i(GD_ERR)) destined to BS
  end if
  if enc_GNOT(JOIN_APRV) from any g ∈ V_GD and hop(s,g)=1
    neighbor_dominator(s)←g
  end if

for each g ∈ V_GD
  if enc_i(JOIN_REQ) from any s ∈ V_Os and s ∈ Os_g
    send enc_G(JOIN_APRV)
```

```
        edge(s,g)
        sub-ordinate(g)←s
    end if
if enc_{iNOT}(JOIN_REQ) from any s∈V_{Os}
    mediator(g)←s
    end if
    if enc_i(GD_ERR) from any s∈V_{Os} and hop(s,g)=1
    report enc_G(ORP_ERR) to BS
    end if
#In case of the BS:
if enc_i(GD_ERR) from any s∈V_{Os} and enc_G(ORP_ERR)
from any g∈V_{GD}
    if same id of s, issue command: Adopter_GD(s)←g
```

Fig. 8. Pseudo Code for Secure Clustering Algorithm

All the groups of sensors could be deployed at a time or more groups could be deployed later based on demand. If it is needed, some sensors in a group could be deployed later. During the offline key pre-distribution, all the nodes are assigned the keys but all the nodes might not be deployed. When any of those remaining nodes is newly deployed, it follows the procedure of joining a group. If authorized by the access list of GD, it joins the group. Otherwise, GD forwards the id of this sensor to BS. BS informs GD about the individual key of that Os if it is a legitimate node. If authenticated by BS, GD generates a new group key and encrypts the new group key with the newly added node's individual key and sends it to that particular Os. All other nodes in the group know about the change of group key by a local broadcast by the GD of that group. In this case, the previous group key is used for encrypting the new group key. For leaving a group or cluster, the node simply leaves a message to inform the GD which in turn generates a new group key and multicasts it within the group members.

## V. PERFORMANCE ANALYSIS AND COMPARISON

We form a WCDS to cover almost all of the nodes in the network with minimum effort. The offline rank assignment reduces the burden of executing resource-hungry operations to form clusters like other clustering mechanisms. As shown in equation (1), WCDS requires less number (or equal to) of dominating nodes to cover the whole network than that of a CDS requires. Depending on the requirements we can increase or decrease the value of η (the expected degree of a GD in a group). In ideal case, the size of the dominating set created in our approach could be obtained by,

$$\text{Size of Dominating Set} = \frac{\text{Number of vertices in the Graph}}{\eta + 1}$$
$$= \frac{\text{Number of vertices in the Graph}}{\Delta(GD) + 1} \quad (2)$$

In our experiment, we generate random graphs of 20-200 and 40-200 nodes with expected average degree 6 and 12 respectively. To simulate the structure of the sensor network, we place the vertices randomly over a 2-D rectangular plane. The network size and density is set by changing the number of vertices and transmission ranges of the nodes. Applying our approach and two algorithms (I and II) of [13] we find that our approach generates much smaller number of group dominators or cluster heads. For a large number of sensors it works effectively. Figure 11 shows the size of dominating sets in comparison with that of Algorithm I and Algorithm II of [13]. The major advantage of our approach is the flexibility to set the value of η (expected maximum degree of a GD) according to the requirements for deploying the network.

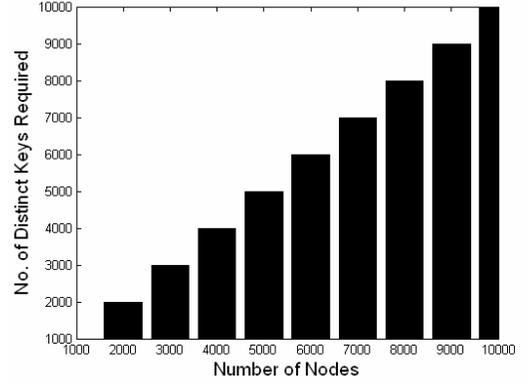

Fig. 9. Number of distinct keys required to support the size of the network

We use distinct group keys for each of the GDs and distinct individual keys for each Os. So, in general case, the number of distinct keys required for our network depends directly on the number of sensors in the whole network (Figure 9).

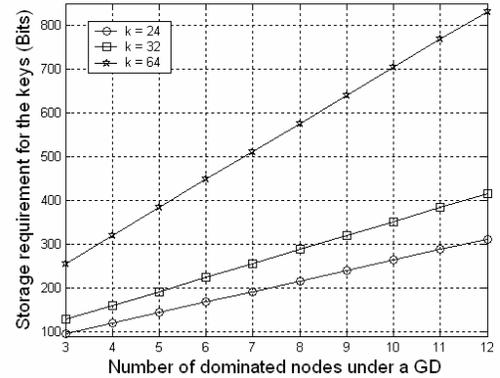

Fig. 10. Storage requirement for a GD for storing the keys for various values of η

Each group dominator (GD) in the network has to remember one group key and all the individual keys of the Oss of that particular group. So, the storage requirement for each GD in number of bits is,

$$\gamma_{GD} = (\eta + 1) \times k \quad (3)$$

and for each Os,

$$\gamma_{Os} = 2 \times k \quad (4)$$

where, η is the number of Oss in that particular group and $k$ is the number of bits required for representing the key. As the value of η increases, the storage load for a GD increases. Hence, the value of η is set according to the requirements or a particular situation at hand. So, if initially we have α number of GDs and β number of Oss, the network wide storage usage for storing the keys is,

$$\Gamma_{network-wide} = \alpha \times ((\eta+1)\times k) + \beta \times (2\times k)$$
$$= k \times (\alpha \times (\eta+1) + 2 \times \beta) \qquad (5)$$

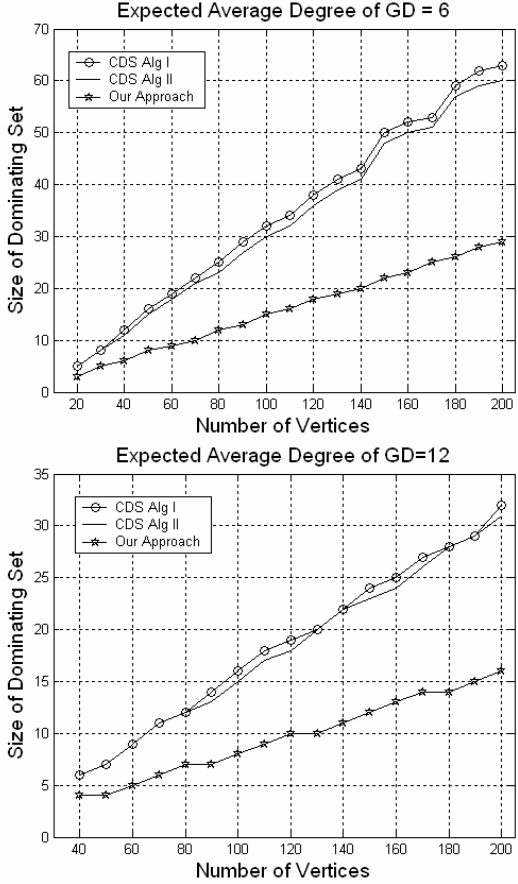

Fig. 11. Number of Vertices versus size of the dominating set when expected average degree 6 and 12

After formation of clusters within the network, the mediators are used for communication among clusters. From the higher level view, we could consider the clusters (or groups) as nodes in a random graph G=(n, p), where n is the number of nodes (i.e. clusters in our case) for which the probability that an edge (i. e. communication link via mediator) exists between two nodes is p. p=0 when there is no edge and p=1 when the graph is fully connected. According to Erdös and Rényi [14], for monotone properties, there exists a value of p such that the property moves from "nonexistent" to "certainly true" in a very large random graph. The function defining p is called the threshold function of a property. Given a desired probability $P_c$ for graph connectivity, the threshold function p is defined by,

$$P_c = \lim_{n \to \infty} P_r[G(n,p) \text{ is connected}] = e^{e^{-c}}$$

$$\text{where, } p = \frac{\ln(n) - \ln(-\ln(P_c))}{n}$$

Let, p be the probability that an edge (communication link via mediator) exists between two GDs of two clusters, n be the number of nodes (i.e. clusters/groups in the entire network in this case), and d be the expected degree of each GD, then,

$$d = p \times (n-1)$$
$$= \frac{(n-1)(\ln(n) - \ln(-\ln(P_c)))}{n} \qquad (6)$$

Figure 12 illustrates the plot of the expected degree of a node d, as a function of the network size n (i.e. here the number of clusters or groups), for various values of $P_c$. The figure shows that the expected degree of a GD needs to be increased by two to increase the probability that a random graph is connected by one order. Moreover, the curves of the plot are almost flat when n is large, indicating that the size of the network has insignificant impact on the expected degree of a node (here, clusters) required to have a connected graph.

In our approach, the sensors could be added later on rather deploying all of them at a time. Sometimes the entire terrain information and deployment diagram could be available (consider a battlefield scenario where the sensors are deployed prior to the enemy forces' invasion). In this case, the extra sensors could be deployed within the range of its appropriate group or cluster. If the sensors are deployed randomly, in the worst case, all the extra or newly added sensors will not be within the range of their intended group dominator and even no other GD could be available in their surroundings. Hence, in the worst case, all the newly added sensors would be included in the dominating set which would increase the size of the dominating set. Still it could be less than the number of dominators needed in case of a connected dominating set (CDS) when the network size is very large.

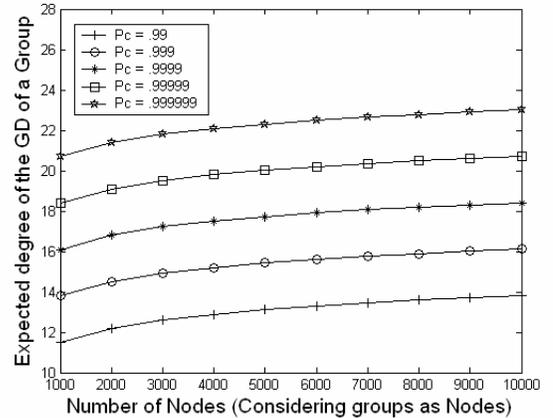

Fig. 12. Given a connectivity probability, expected degree of a GD from the high level view

Our scheme ensures that, each of the GDs and the corresponding Oss could directly form the groups (i. e. clusters) maintaining the security of the network from the bootstrapping state. As encryption is used for message-transmission within the network from the very beginning of the network formation, our scheme could successfully defend Hello Flood Attack [15] and most of other attacks in wireless sensor networks [16]. Again, as each node carries distinct individual and group keys, compromising one node affects only one link in the network while other links remain safe from the attacks by the adversaries. If the group key of a particular group is compromised, still the adversary needs valid individual keys of the Oss for decrypting the information sent from an Os. In case of the compromise of a GD, the base station gets involved for revoking the keys and even in this case, only one group is affected while others could still operate.

As the group dominators rule over all other sensors in the group for data transmission, the dominators could require more energy, processing and storage power. For this, a set of sensors with greater resources could be considered as dominators. In our approach we kept the number of cluster heads small; hence, it could reduce the cost in comparison with a network which requires a large number of cluster heads. To avoid traffic concentration on a few cluster heads, the cluster size should be evenly distributed among the cluster heads. Our approach could perform well in this case. To reduce inter-cluster-head traffic, the number of clusters should be controlled and in our scheme, as the number of cluster heads is relatively less, the number of clusters is relatively less. Keeping the size of the dominating set (i. e. number of cluster heads) to a minimum also helps for better security in the network because, there is comparatively small number of entry-paths (to the base station) for injecting false report by the adversaries and each dominator in the set could check the validity of the reports sent from the subordinate sensors before forwarding those to the base station. For this the GD could set a particular value τ, which is the number of sensors in a particular group/cluster that should send the same report to the GD to convince that the report is true. The $GD_{os}$ that could be formed in the network does not have any subordinate sensors, hence it could easily carry an extra group key in its memory. Basically in that case, only the rank of the Os changes but it does not incur any significant load on it.

Our approach has some limitations. If a sensor network is deployed via random scattering (e.g. from an airplane), the sensors could be well-scattered even if one group is released at a time (the worst case as mentioned earlier) and the nodes of the same group could be out of the communication ranges of each other after deployment. Even if the nodes are deployed by hand, the large number of nodes involved in DSN makes it costly to predetermine the location of every individual node. The re-keying feature ensures robust security as with each addition of a new sensor, the group key is renewed but, it could be resource-exhaustive for the resource-constrained sensors. In such a case, the key renewal mechanism could be omitted. However, for military networks as security is the major issue, we could consider a slight increase of the usage of the resources in the sensors.

## VI. Conclusions and Future Works

This paper has presented an efficient approach for secure clustering in distributed sensor networks based on key-predistribution and prior rank assignments. We have dealt with only the clustering phase of a distributed sensor network. There is still a lot of scope to extend the work further. In future, we will try to mitigate or remove the limitations of our approach and will deal with secure routing and an efficient method to prevent Denial-of-Service (DoS) attacks in distributed sensor networks using our approach.